# ZnO/LSMO Nanocomposites for Energy Harvesting

Robert Kinner[1$], Abdul-Majeed Azad[1*], S. Gopalan[2], S. Gallapudi[2], Menka Jain[3]
[1]Department of Chemical Engineering, The University of Toledo, Ohio, USA 43606.
[2]Department of Physics, Oakland University, Rochester, Michigan, USA 48309.
[3]Department of Physics, University of Connecticut, Storrs, Connecticut, USA 06269.

[$]Current address: FirstPower Group LLC, Twinsburg, Ohio, USA 44087.

## Abstract

The composites of strontium-doped lanthanum manganite (LSMO) with zinc oxide (ZnO) are candidate materials for energy harvesting by virtue of their magnetic and piezoelectric characteristics. They could be used to harvest energy from stray sources, such as the vibrations and electromagnetic noise from transformers and compressors within electrical grid power stations to power small diagnostic sensors, among other applications. The LSMO/ZnO nanocomposites were made by: (i) milling the two bulk powders and, (ii) a wet chemical process which resulted in core-shell structures. The electrical, piezoelectric, and magnetoelectric properties showed strong dependence on the fabrication method. Growth of ZnO nanopillars on the particulate core of LSMO surface appears to have improved the piezoelectric properties. Moreover, the chemical bath deposition process can be easily modified to incorporate dopants to augment these properties further.

---

[*] Corresponding author; abdul-majeed.azad@utoledo.edu;   +1 419-530-8103

# 1. Introduction

With the long term uncertainty of fossil fuel resources and the environmental repercussion such as the global climate change as a result of their continued usage, our planet is arguably facing an impending energy crisis.

The concerns about the planet getting hotter and energy consumption exceeding production capacity are merely the symptoms of the underlying problem that the world is not utilizing energy properly. The increasing atmospheric concentration of carbon dioxide is a prime example that energy is being over-utilized from fossil fuel sources, and underutilized from other sources. Since green vegetation and plants utilize carbon dioxide to store sun's energy in the form of carbohydrates and lipids, biofuels are being pursued as possible alternatives to fossil fuels. However, in this case, the primary problem lies with the socio-economic burden associated with the production of biofuel in volumes needed to meet the current and future demands. Some experts argue that an enthusiastic but unwise pursuit of this route may alter the landscape of animal grazing and agricultural zones irreversibly worldwide. In this respect, fossil fuels derived from the accumulation of decaying organic matter over thousands of years have had a bit of a head start. Arguably, the wind and solar energy technologies are useful, but, their dependence on day-to-day weather patterns makes them unreliable as a primary energy source. In addition, they require considerable capital investment with a long buyback period.

Hence, new sources of ambient energy have been identified and new methods of energy harvesting are being explored. Common forms of harvested energy are vibrational, electromagnetic radiations, solar, radiofrequency and thermal, to name a few.

Figure 1 shows a comparative speculation of various form of energy for 5 years up to 2014. There is more demand for the electromagnetic energy harvesting devices than for the other forms combined[1]. This is likely due to the increasing number of sensor networks being integrated with the electrical power transmission grid, anticipated by the Department of Energy by 2030[2].

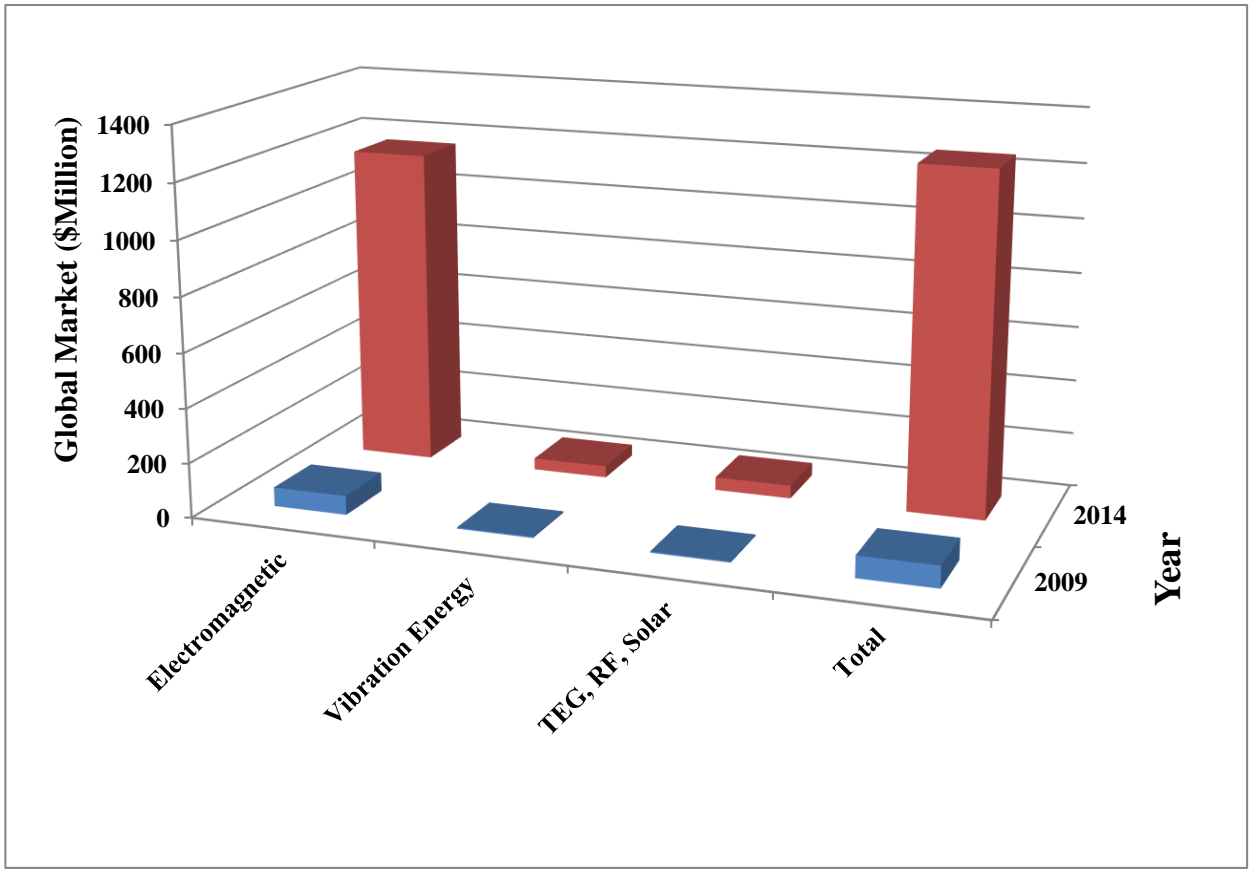

**Figure 1**: Global energy harvesting market speculations for 2014[1].

Through the use of scientific rationale and incorporation of smart ceramics, energy harvesting devices could be developed to convert energy from various benign - and sometimes stray – sources. While the output of such a device might not be large enough to power a home or run a laptop, it may nevertheless be enough to power a small device, such as a sensor. Thus, by tapping into these unused ambient sources, one can provide sustainable energy for small, low-power devices, and effectively increase the feasibility of alternative and green energy sources to penetrate commercial markets.

Magnetoelectric materials possessing both piezoelectric and magnetostriction attributes constitute one such smart system. They find wide ranging applications, such as sensors, data storage devices and for energy harvesting. When exposed to a magnetic field, the magnetic domains of the magnetostrictive components attempt to align and/or orient along direction of the magnetic field lines. This induces a strain, which is absorbed by the piezoelectric element, thus creating a voltage – commensurate with the magnitude of the induced strain. These properties are either inherent in a single material or can exist as a net effect in a composite of multiple materials. The latter situation allows the possibility of engineering the materials for specific purpose.

There has been significant work on particulate magnetoelectric materials by way of coupling piezoelectric material such as barium titanate with magnetic cobalt ferrite particles made via traditional and/or sol-gel technique. In the case of sol-gel, the magnetic phase is encapsulated by the piezoelectric material, and displayed higher magnetoelectric effect[3-6].

With a focus on energy harvesting, recently wurztite-structured ZnO-based piezoelectrics with crystal lattice tuned to respond to vibration have also been pursued[7-8]. Chemical bath deposition processes, utilizing the degradation of hexamethylene tetra amine (HTMA) have been shown to form ZnO nanorods with the desired hexagonal crystal motif[9-12].

With these attributes in mind, the goal of present investigation was to examine the coupling of zinc oxide (ZnO) and perovskite-structured strontium-doped lanthanum manganite (LSMO) particles made in the core-shell configuration, and compare with the composites made via traditional processing methods, in terms of their magnetoelectric response. Using LSMO-based magnetostrictive particles as nucleation centers, zinc oxide could be deposited on them to yield a core-shell configuration to the resulting composite.

The rationale for employing LSMO is as follows: It acts as the magnetostrictive element in the composite, for its well-known colossal magnetoresistance (CMR) properties[13-14]. LSMO's magnetostrictive characteristic as a result of change in entropy leads to magnetocaloric effect[15-17]. This gives rise to its potential use in targeted drug delivery application by way of hyperthermia[18-19].

## 2. Experimental

The strontium-doped lanthanum manganite (LSMO) was synthesized via organometallic combustion reaction using glycine as the fuel in the presence of ammonium nitrate as the igniter. A typical batch of the perovskite powder was made by dissolving in deionized water 15.146g of lanthanum nitrate hexahydrate, 1.828g strontium nitrate and mixing with 15.77g of manganese nitrate 50% (w/w) solution. All the solutions were transferred to a 1L capacity stainless steel vessel which was covered with a perforated aluminum foil and heated on a hot plate at 80°C under constant stirring to slowly boil off the excess water. The solution eventually formed a viscous gel, which ignited releasing considerable amount of heat. The resulting brownish-black fine powder was extracted, crushed and homogenized in an agate mortar and pestle. It was calcined at 600°C for 2 h, homogenized again and sieved through a -325 mesh screen.

For obtaining a core-shell composite, zinc nanorods were fabricated from an aqueous solution of zinc nitrate, with hexamethylenetetramine (HMTA). For this, 250mg of LSMO powder prepared above was suspended in 300mL deionized water containing 8.97g zinc nitrate hexahydrate (FW = 297.47) and 4.21g HMTA (FW = 140.18). This constituted an equimolar mixture of zinc nitrate and hexamine. The mixture was transferred to an airtight glass vial which was kept in a sonicator (to prevent settling of the particulates) and heated at 70°C for 4 h.

Composites were also fabricated by mixing bulk ZnO and LSMO powders in the weight ratio of 1:1, 3:1, 1:3 and 9:1; the last composition was found to be the most optimal in terms of the magnitude of the magentoelectric coefficient. The mixture was placed in a Nalgene container and ball-milled using zirconia milling media for 4 h in acetone, followed by drying overnight in an air oven under partial vacuum.

The dried powder was pressed into pellets (14 mm diameter and 3 mm thick) using a uniaxial Carver press, and fired at 800°C for 2 h. The opposite faces of the pellets were polished using fine-grit sandpaper. Silver conducting paste was applied evenly and uniformly on both sides of the pellets, which were fired again at 600°C for 2 h to cure the conducting paste which formed ohmic contacts for electrical measurements. The edges of the post-cured pellets were polished using fine-grit sandpaper to ensure uniform dimensions and to prevent short-circuiting.

The piezoelectric and magnetoelectric characterization of the composites was performed by using an experimental set shown in Figure 2. The test setup for determining magnetoelectric coefficient consisted of the following: power supply, Helmholtz coil, multimeter, function generator, lock-in amplifier, permanent magnets, Hall probe, and a desktop PC. The piezoelectric coefficient ($d_{33}$) was measured using a Pennebaker (Model 8000) piezo d33 tester. The piezo tester applies a pressure to the sample which generates a charge, Q. By utilizing a variable capacitor, an AC voltage is obtained from V = Force · Distance ÷ Q, which is measured on a digital voltmeter. The $d_{33}$ coefficient is –Q/Force.

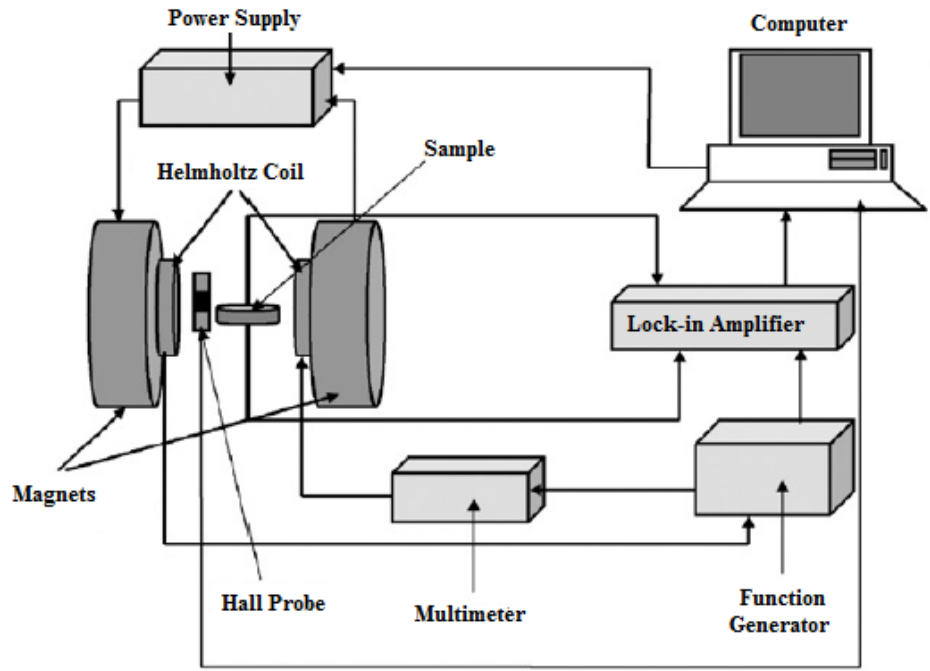

**Figure 2:** The magnetoelectric characterization setup used in this investigation.

## 3. Results and Discussion

The attempts to coat the LSMO particles with ZnO nanorods via deposition in a chemical bath proved successful. As can be discerned from the high magnification SEM images in Figure 3, the LSMO surface was extensively covered with tiny hexagonal ZnO nanopillars.

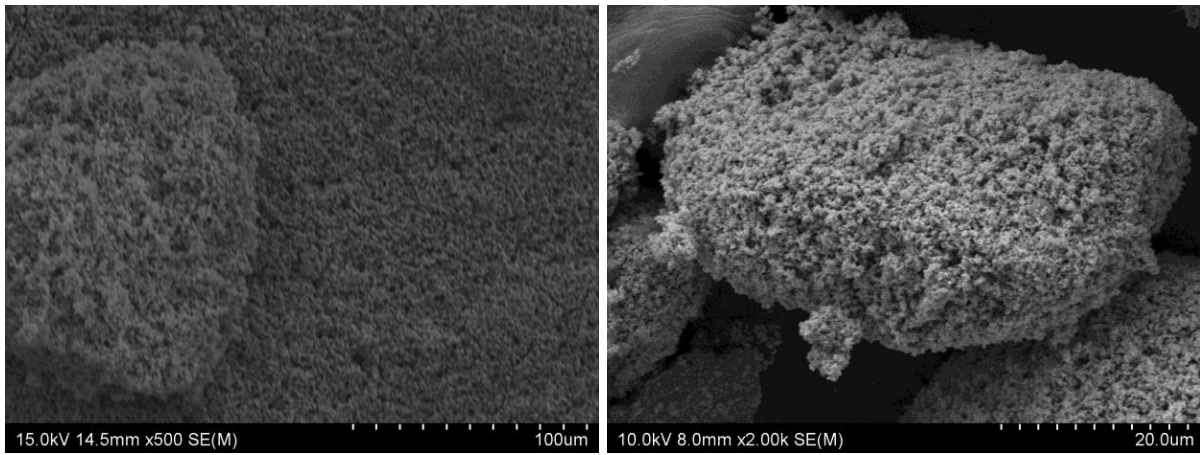

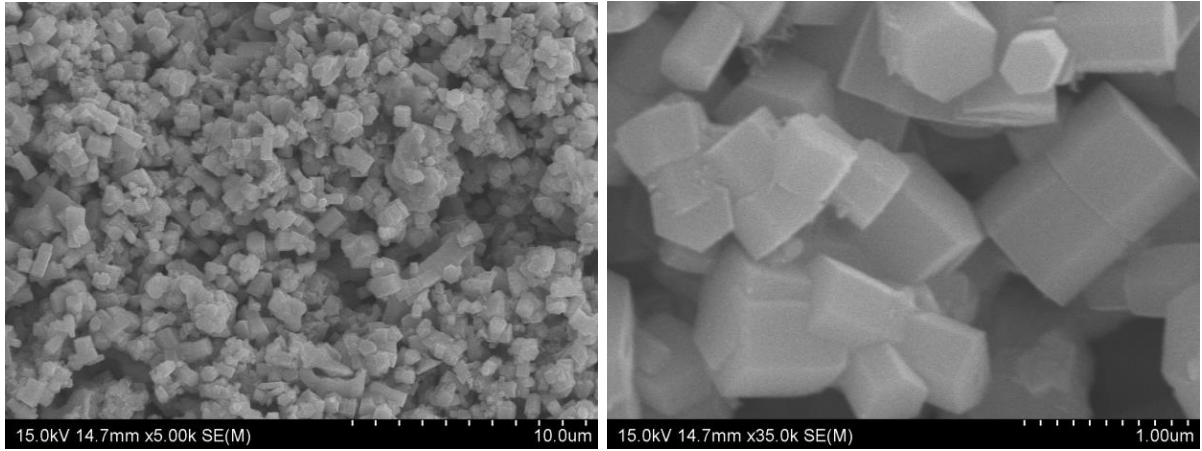

**Figure 3:** SEM images (at different magnifications) showing the growth of ZnO nanopillars on LSMO surface.

The microstructural images collected on the bulk samples are shown in Figure 4. They served as a reference with which the morphology of the core-shell samples was compared.

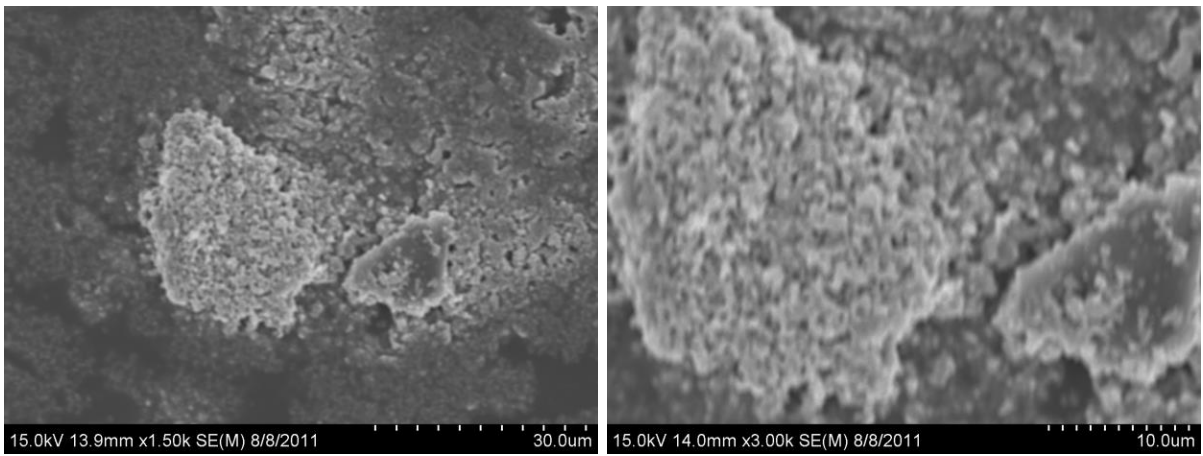

**Figure 4:** SEM images of the bulk powders containing ZnO and LSMO in 9:1 weight ratio.

Subtle changes in the way the composites were fabricated and processed, created profound difference in their electrical characteristics, as can be seen from the data summarized in Table 1.

Table 1: The piezo and magnetoelectric response of LSMO-ZnO composites

| Sample | Resistivity (Ω·cm) | $d_{33}$ (pC/N) | ME Coefficient (µV/cm·Oe) | Power @ 20 Hz (nW/cm·Oe) |
|---|---|---|---|---|
| Bulk | $6.69 \times 10^8$ | 3.4 | 136.9 | $2.5 \times 10^{-7}$ |
| Core-Shell | $6.42 \times 10^4$ | 13.4 | 1.9 | $4.95 \times 10^2$ |

For example, the resistivity of the two samples differed by several orders of magnitude. There are two possible explanations for this phenomenon. One, the LSMO has better conductance, is more evenly distributed throughout the pellet, and therefore, less likely to form agglomerates. Second, zinc oxide loses oxygen during the sintering process, thereby creating oxygen vacancies and resulting in a metallic phase.

It is also interesting to note that the piezoelectric coefficient in the case of core-shell structure was about four times that of the bulk sample. Also, the coefficient $d_{33}$ of the ZnO-LSMO core-shell composite is significantly higher than either the bulk ($d_{33}$ = 9.9) or properly aligned ($d_{33}$ = 12.4) pure ZnO[20]. It is likely that electromagnetic coupling optimizes the orientation of the zinc oxide.

Azad and Jain[21] have measured the resistance and magnetorestirction of the sintered pellets made from bulk LSMO as well as electrospun nanofibers, under magnetic field between zero and 7 Tesla. As expected, the sample made from electrospun nanofibers had lower resistance; it also exhibited higher magnetorestriction (%MR). The inflection point in the MR curve appeared at slightly higher temperature (~310K) in the case of nanofibers than the bulk LSMO samples (~273K). This is shown in Figure 5.

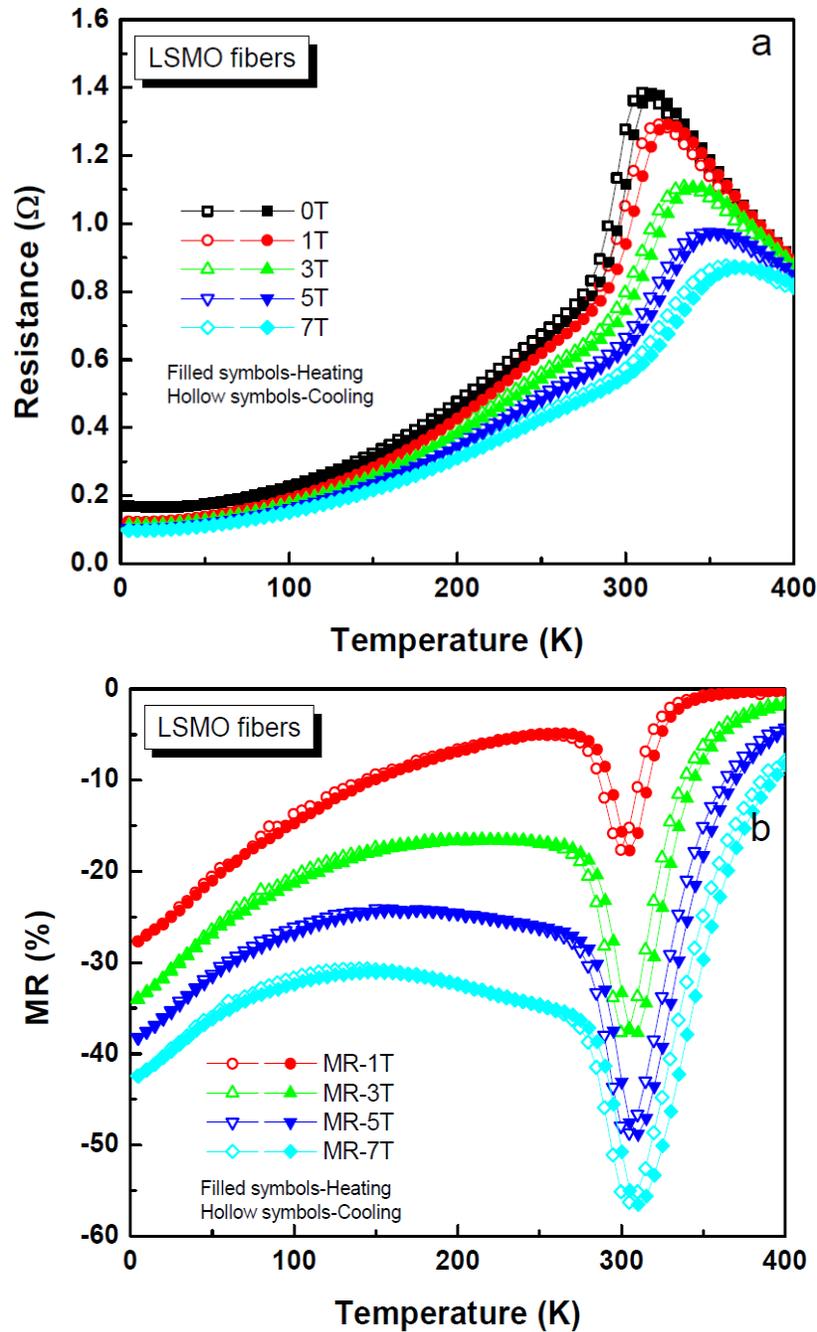

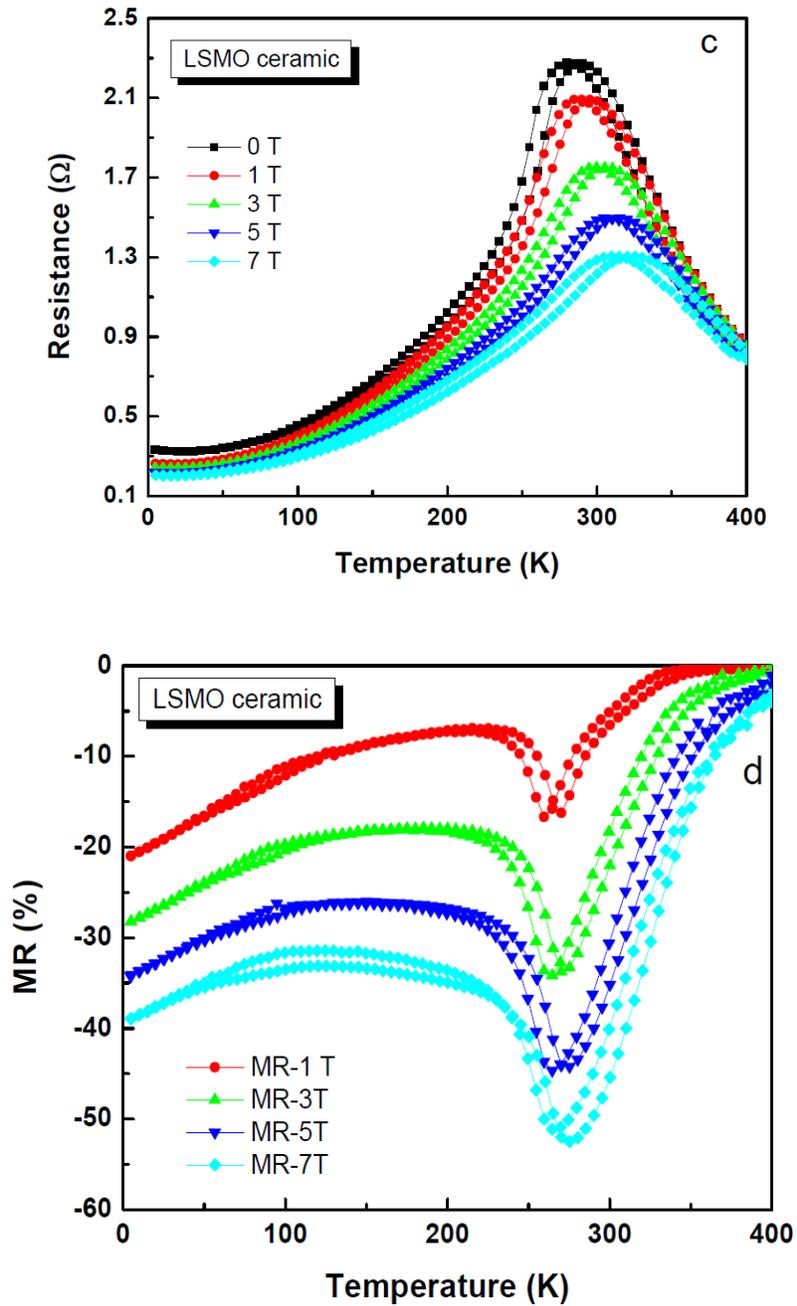

**Figure 5:** Temperature dependence of the resistance and percent magnetorestriction in the sintered pellets made from nanofibrillar (a and b) and bulk (c and d) LSMO samples, at different magnetic field strengths[21].

Magnetoelectric (ME) curves were generated for the bulk as well as the core-shell sample. These are shown in Figure 6 and 7, respectively.

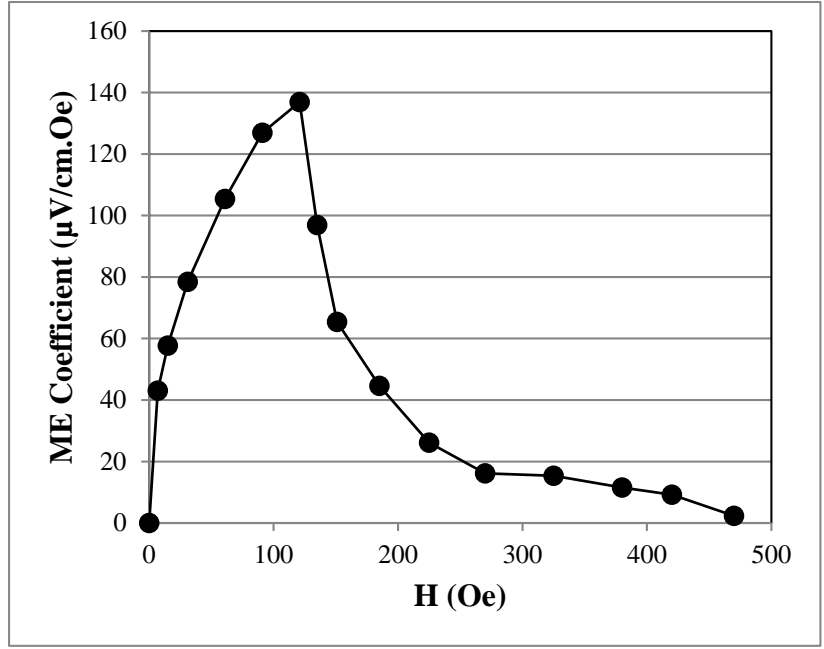

**Figure 6:** Variation of the ME coefficient with the imposed magnetic field strength for the bulk composite.

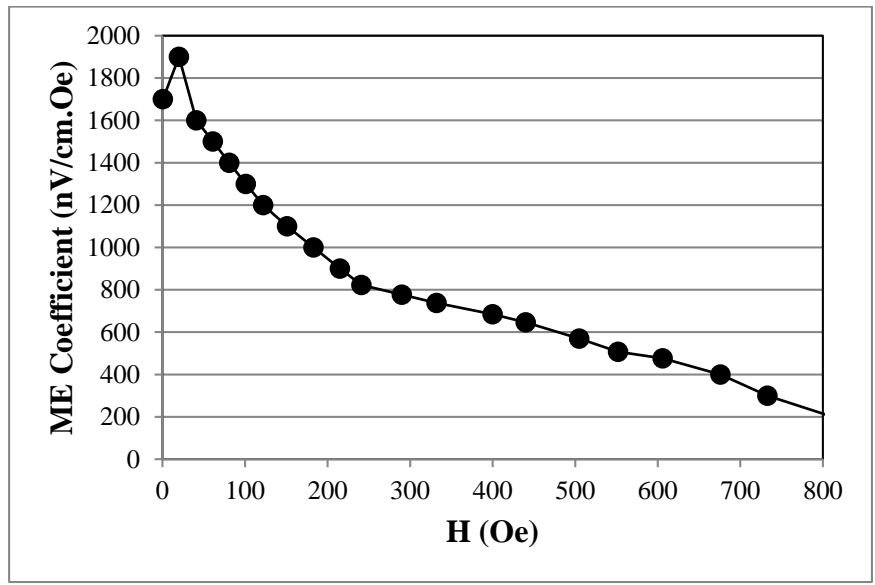

**Figure 7:** Variation of the ME coefficient with the imposed magnetic field strength for the core-shell composite.

As can be seen, the voltage generated across the core-shell sample was very low, which is expected, since the resistance across the pellet was quite low (Table 1). In the case of the bulk sample, the voltage generated was much higher, leading to a commensurately high ME coefficient. Therefore, in cases where high voltage readouts are all that is desired, the bulk sample appears superior to its core-shell counterpart. For such sensor applications, it would seem more feasible to produce bulk composites using perovskite piezoelectric materials in the barium- strontium and/or lead titanate family, since they have a higher resistivity and much larger piezoelectric coefficients.

Since the goal of this work was to explore materials for their use in energy harvesting, the ME coefficient (which is quantified in terms of voltage per unit length per unit magnetic field), may not be the right parameter to rate a material system for this application. For energy harvesting, it is the power output that is more appropriate. The power output of the samples may be calculated by applying Joule's first law to the ME data, by using the relationship, $\text{Power} = \frac{V^2}{R}$.

Thus, by focusing on the power generation, instead of voltage, the core-shell composite proves its superiority, as seen from Figure 8.

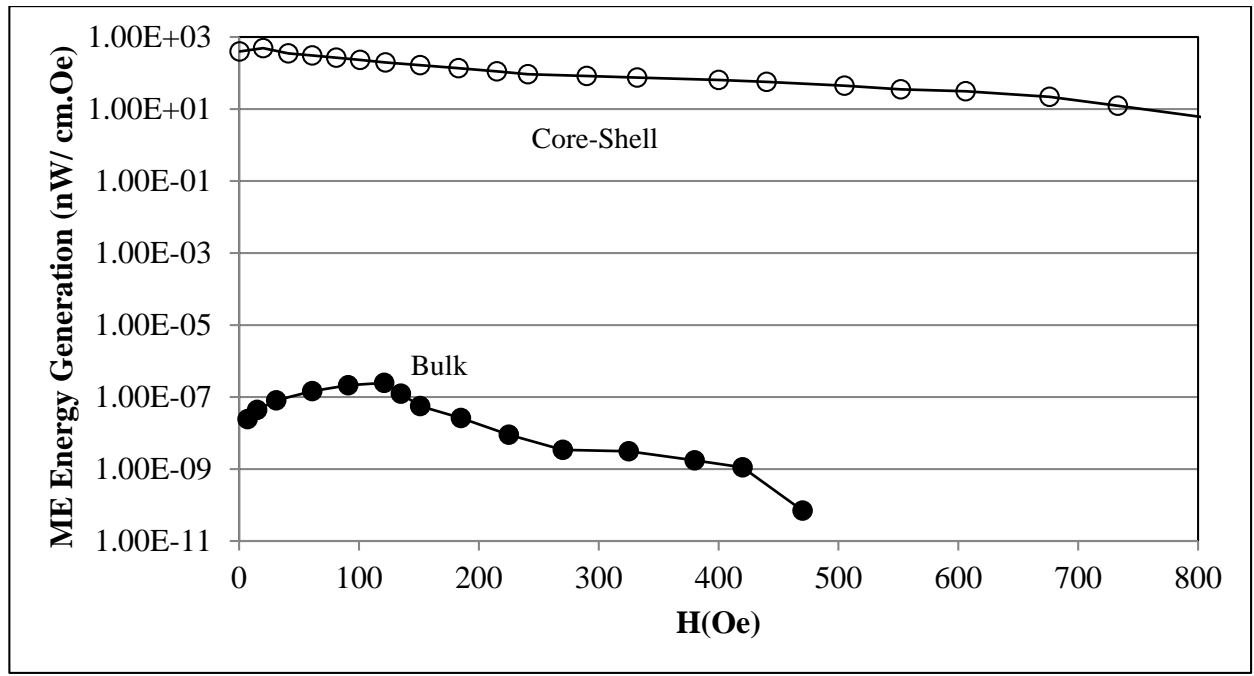

**Figure 8:** Comparison of the power generated by the two composites from 20Hz oscillating magnetic field.

The computed ME power generated by the bulk (9:1 w/w) and core-shell ZnO-LSMO composite samples, as a function of applied magnetic field at low frequencies is shown in Figure 9. The range of test frequencies and the magnetic field was chosen to simulate conditions that are typically encountered in a power transformer. As can be seen, the power outputs were quite small (<1 nW) in both the cases.

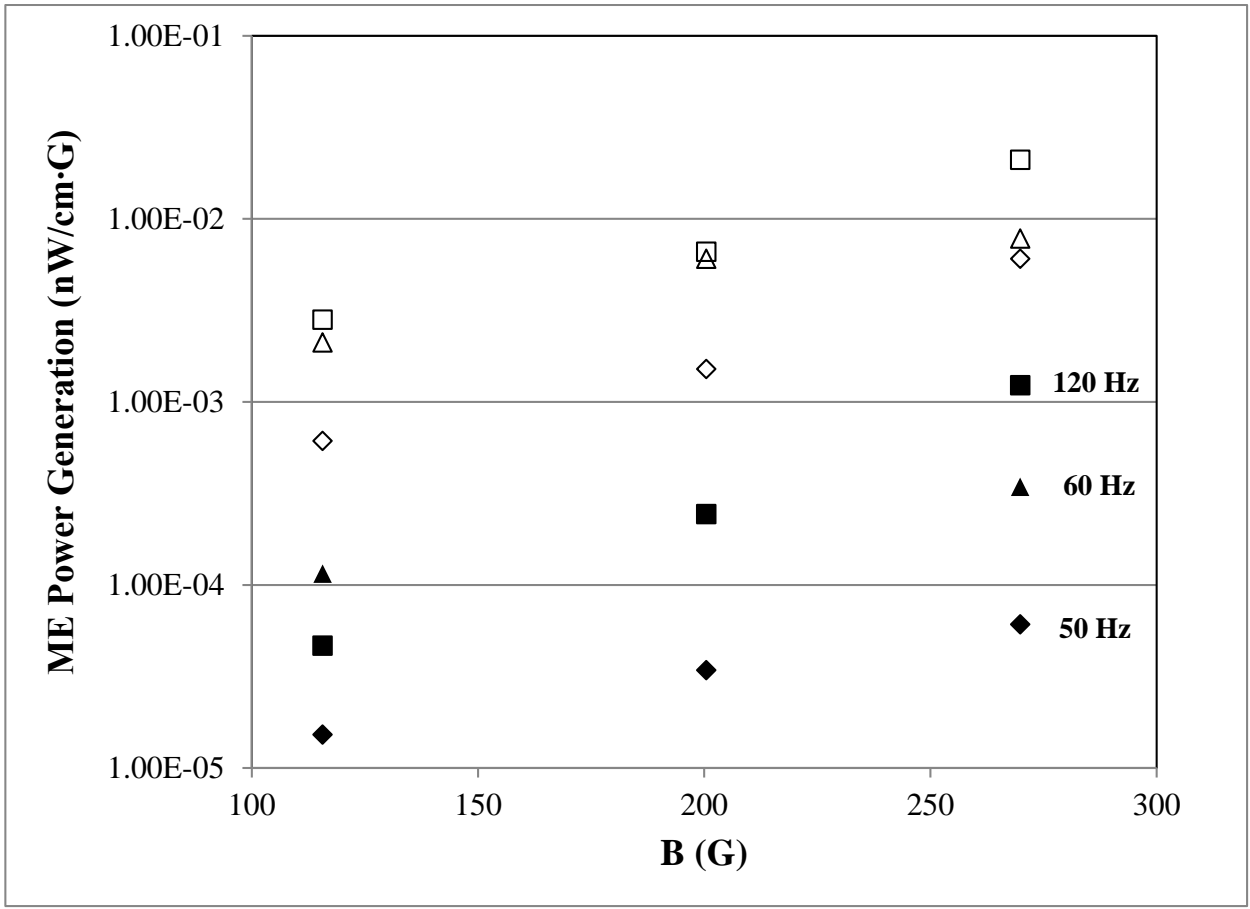

**Figure 9:** Dependence of ME power generated by the ZnO-LSMO composites on the magnetic field strength (B, measured in Gauss) at different frequencies. The open symbols represent the core-shell samples and the solid symbols are for the bulk samples.

Higher frequencies were also used to further evaluate the performance of these composites and the results are shown in Figure 10. As frequency increased into the radiofrequency (RF) regime, the power output of the core-shell composite also increased monotonically, whereas that for the bulk composite remained almost flat over the entire range.

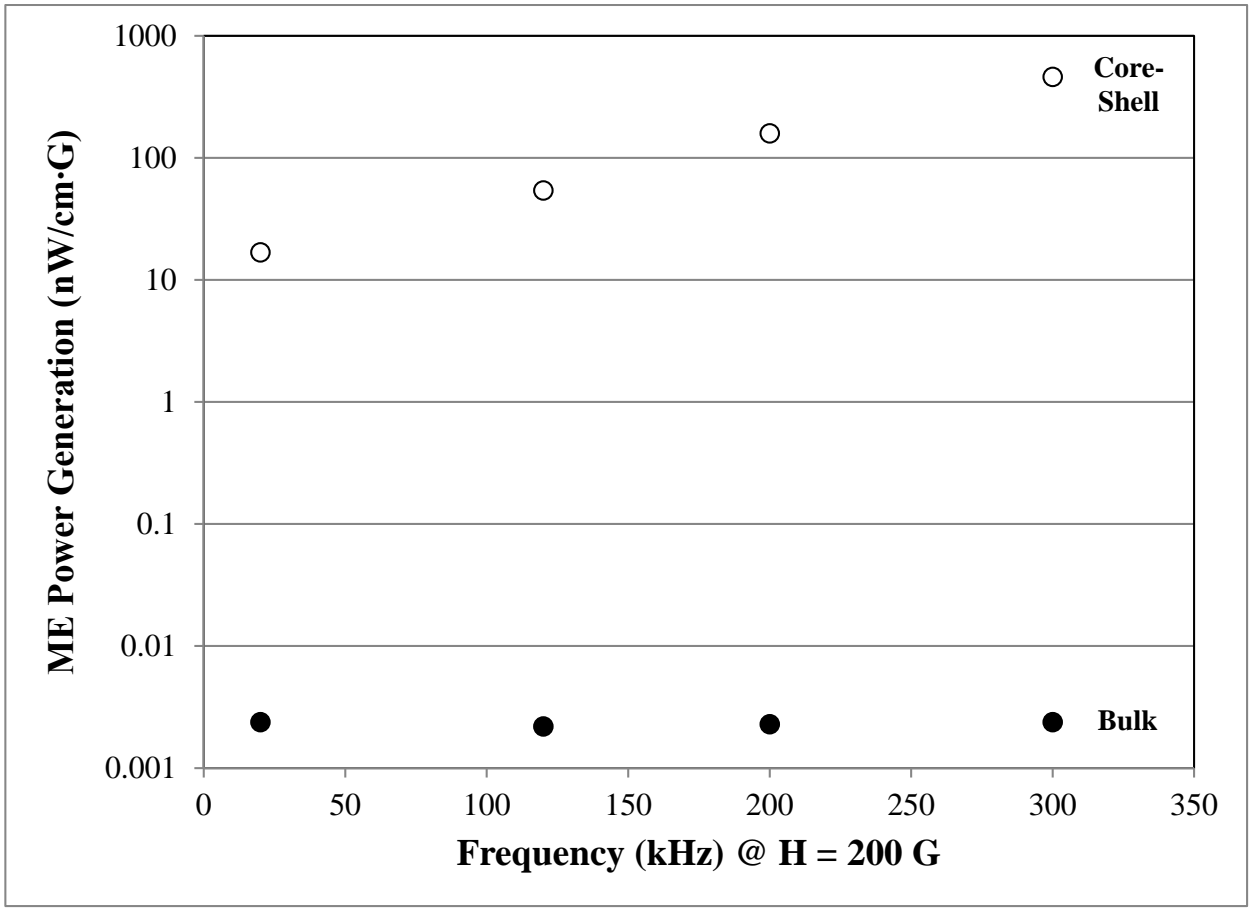

**Figure 10:** Dependence of ME power generated by the core-shell (open circles) and bulk (solid circles) LSMO-ZnO composites on the imposed frequency at a constant magnetic field strength of 200 Gauss.

The core-shell composites were finally tested at a frequency of 1 MHz, which is the highest frequency the test setup could produce, before the signal became distorted. The results are shown in Figure 11. At this high frequency, the ME power was about two orders of magnitude higher and increased linearly with the applied field.

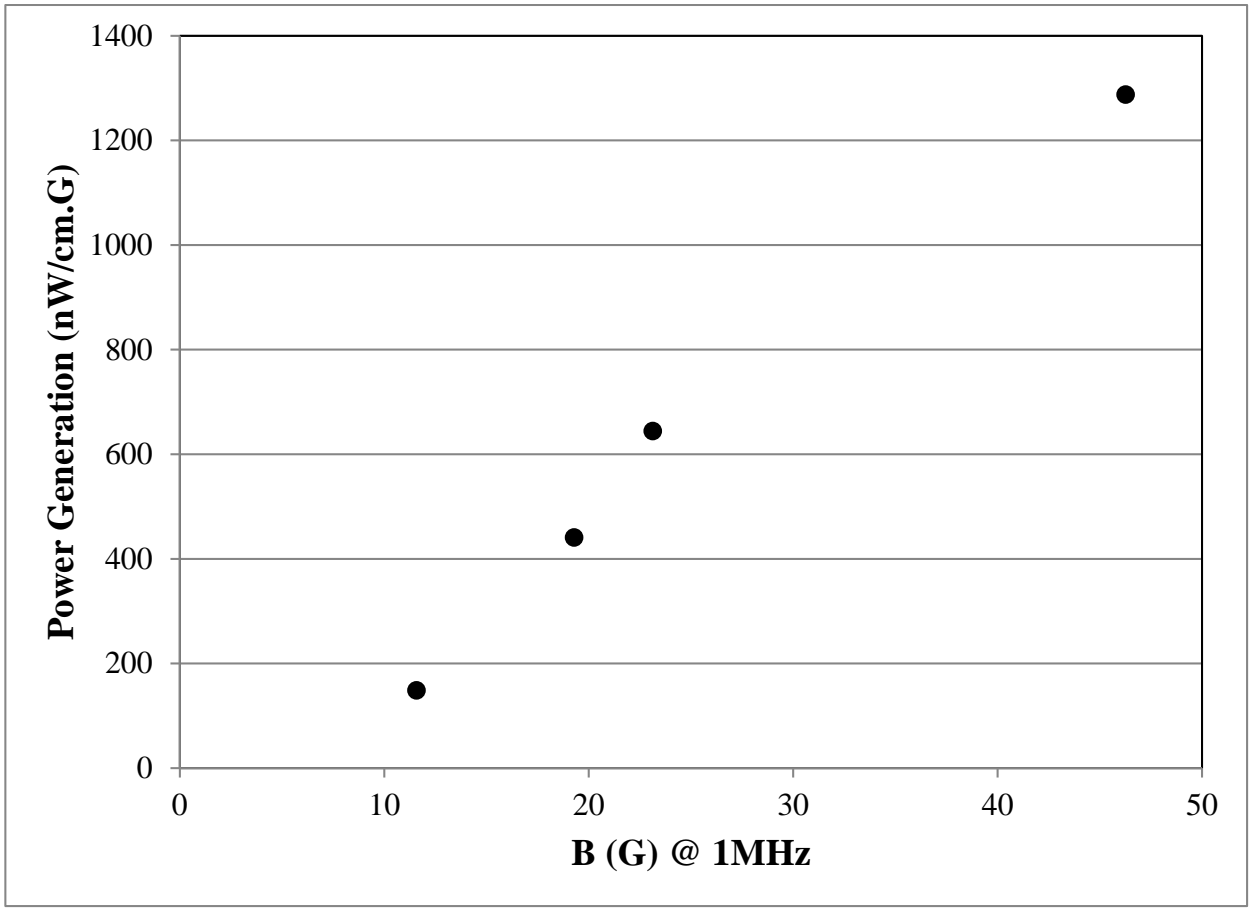

**Figure 11:** Dependence of the ME power generated by the core-shell composite on the magnetic field strength at a constant frequency of 1 MHz.

With the intuition that some compositional modification of the ZnO shell might help to increase the power output, 2.5 wt% $V_2O_5$-doped ZnO shells (hereafter referred to as VZnO) were synthesized on the LSMO core by co-deposition in a slightly modified chemical bath process described in the previous section. The chemical bath deposition process was shown to have the ability to form VZnO homogeneously. The powder x-ray diffraction pattern of the so-synthesized VZnO particles is shown in Figure 12 which appears similar to what has been reported by Yang et al. [13].

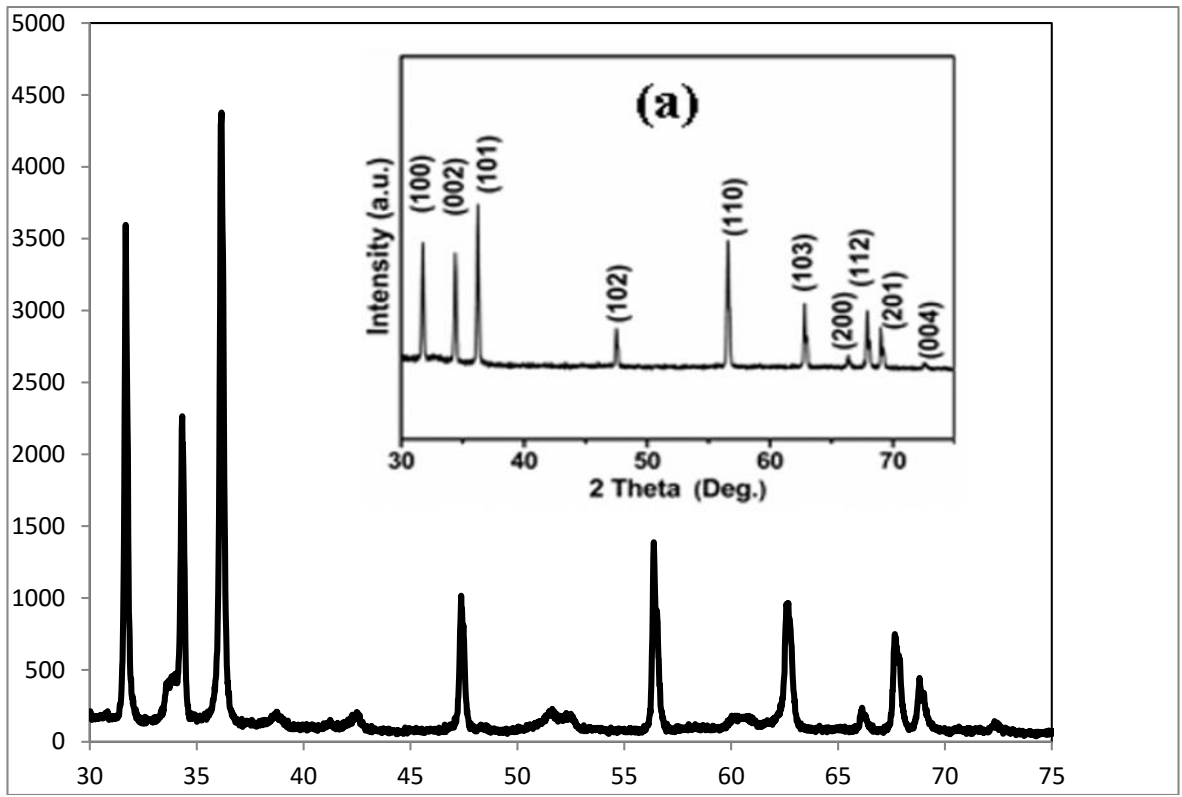

**Figure 12:** XRD pattern of VZnO particles prepared by co-deposition process in a chemical bath; inset (a) is the XRD signature of VZnO from the literature [13].

SEM images of the $V_2O_5$-modified ZnO powder shown in Figure 13 indicate the presence of mixed morphologies; the ZnO nanopillars appear to be surrounded by "flower-like" regions that are rich in vanadium. It is likely that the vanadate ion in the precursor solution were able to undergo complexation, leading to the formation of vanadium-rich "flower-like" regions.

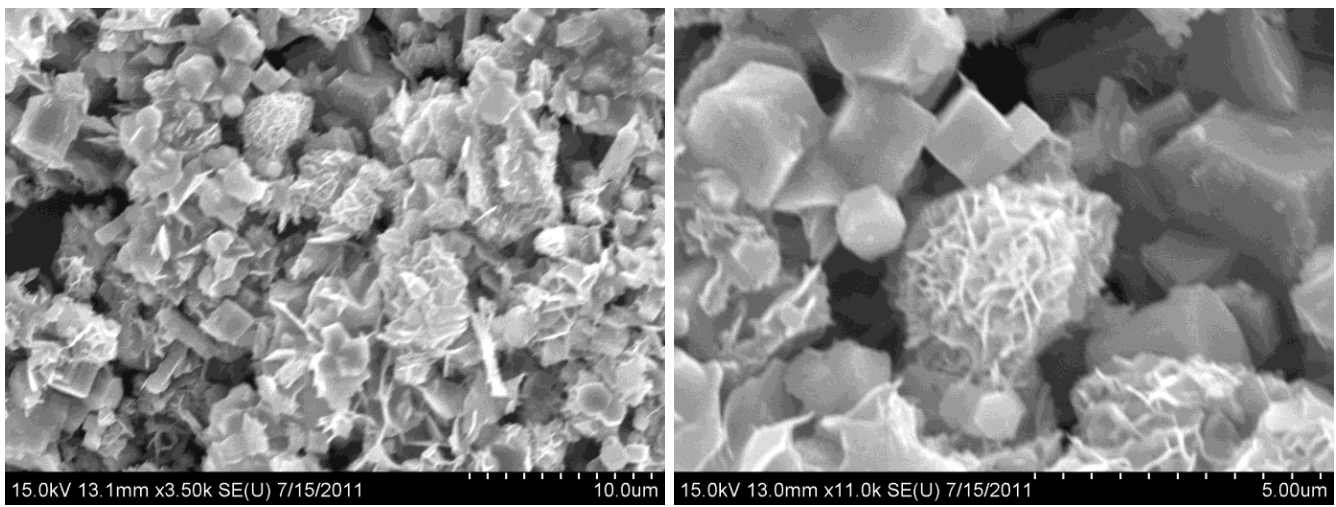

**Figure 13:** SEM images of the VZnO particles at different magnifications.

When grown on the surface of LSMO core, the surface morphology is retained as seen from the SEM images in Figure 14.

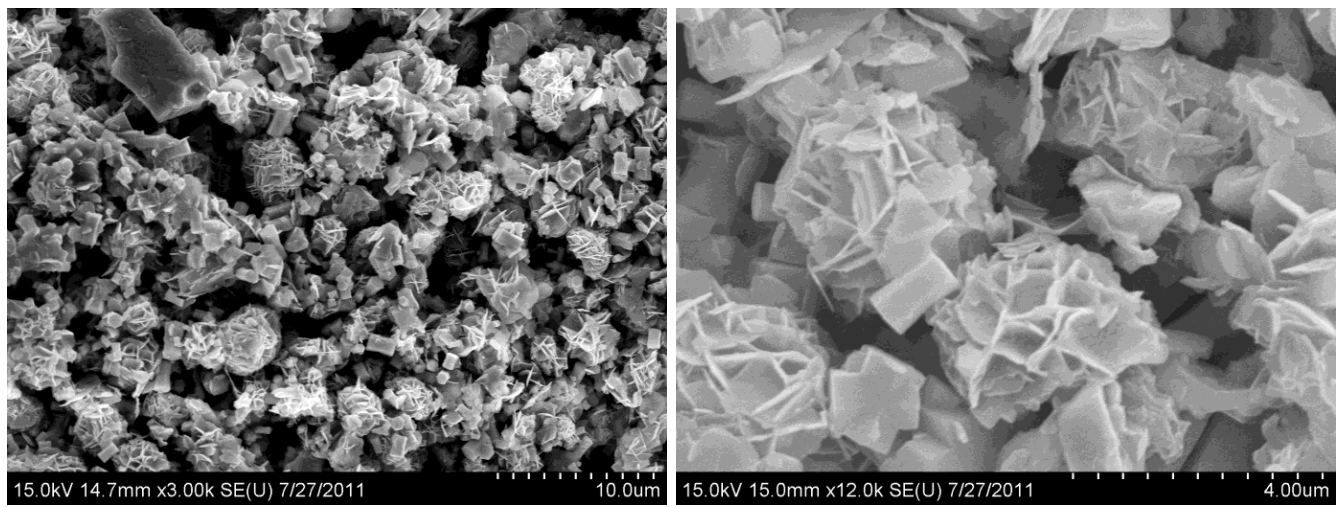

**Figure 14:** SEM images of the VZnO nanostructure grown on LSMO particles at different magnifications.

The power generated by the VZnO-LSMO nanocomposite is compared with other systems in Figure 15. As can be seen, the core-shell samples performed the best; pure ZnO pellet showed no significant power generation at its own. It is therefore, clear that the coupling with LSMO creates the magnetoelectric effect and increases the power output; the core-shell configuration offers further improvement. The vanadium doping further accentuates the ME effect several fold.

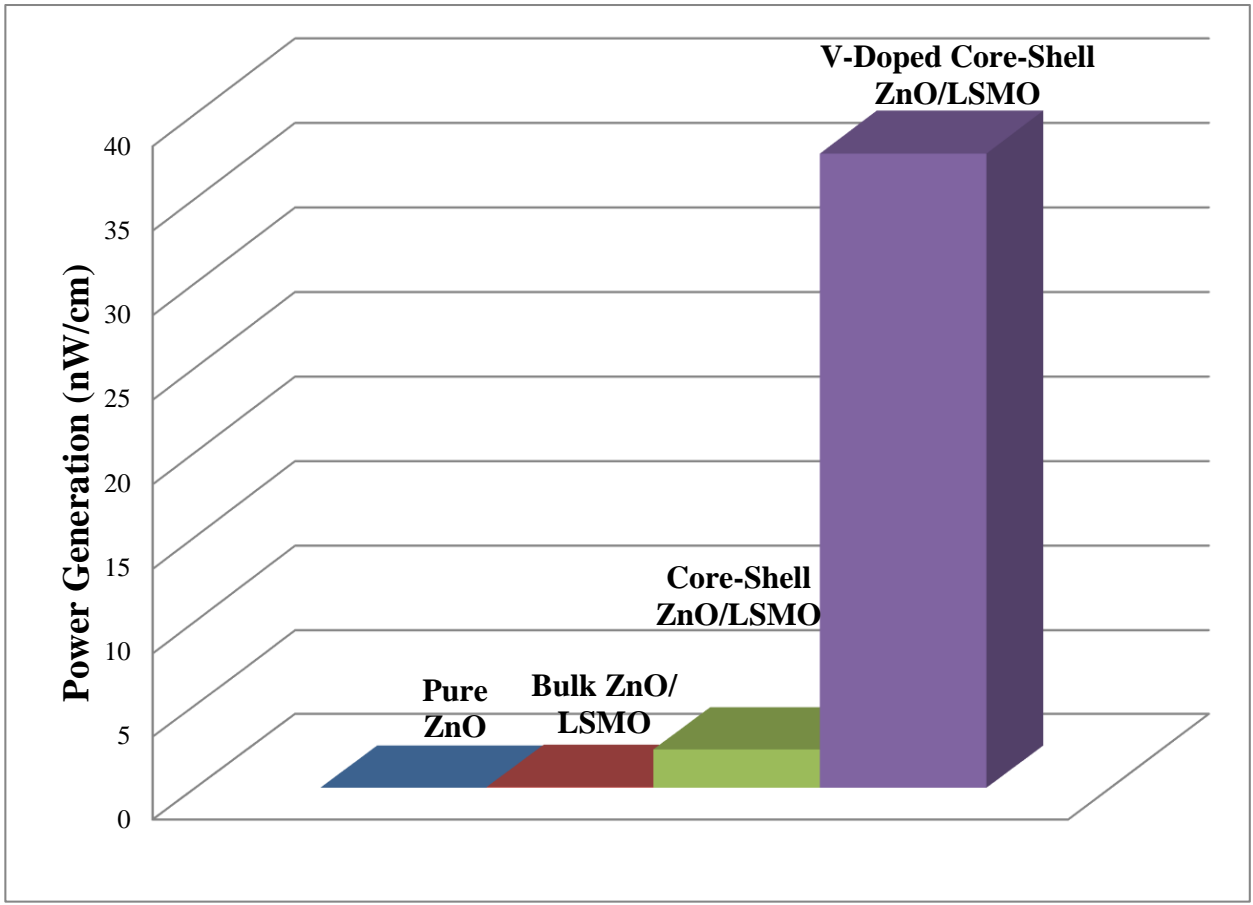

**Figure 15:** Comparison of the power generated by different formulations.

## 4. Conclusions

The creation of magnetoelectric formulations comprising composites of LSMO and ZnO through the use of chemical bath deposition has successfully created nanostructured core-shell powder with superior energy generation properties to those of the bulk composites of same composition manufactured via traditional ceramic routes. This is thought to be the result of lower impedance across the sensor by optimizing the LSMO portion for improving electrical conductivity. The growth of ZnO as nanopillars on top of the particulate core surfaces has been shown to improve their piezoelectric properties. This could be used to harvest energy from stray sources, such as the vibrations and electromagnetic noise from transformers and compressors within electrical grid power stations to power small, diagnostic sensors, among other applications. Use of the bulk powders, provides a greater voltage signal, which could be used in sensor application, where current is not significant.